# Isaac Thuret: celebrated craftsman denied intellectual credit

*Richard de Grijs*
Department of Physics and Astronomy, Macquarie University, Balaclava Road, Sydney, NSW 2109, Australia

*The Scientific Revolution sweeping through seventeenth-century Europe led to unprecedented intellectual and scientific insights and high-profile technological developments. Combined with a significant worldwide increase in naval commerce, solving the intractable 'longitude problem' became an ever more urgent requirement for the continent's main sea-faring nations. Christiaan Huygens, one of the brightest contemporary natural philosophers, established a fruitful professional collaboration with the Parisian master clockmaker Isaac Thuret. Their joint efforts eventually led to the construction of the first accurate, spring-driven watches. Despite clear evidence of Thuret's intellectual contributions, but in the absence of a robust intellectual property rights framework, Huygens insisted on claiming the invention's sole ownership. Thuret, the celebrated craftsman who had contributed crucial advice to realize the novel watch design, was thus forever—and wholly undeservedly—marked as the 'invisible technician.'*

## *A century of scientific progress*

Seventeenth-century Europe witnessed a re-awakening of critical thought. Humanist developments, which appeared from the late Middle Ages onward, and renewed interest in both classical Greek philosophy and pursuits of 'the truth' during the Renaissance led to the Age of Enlightenment by the end of the century. Never before, nor ever since, were new scientific insights so readily attainable.

Developments in science and technology set the tone for the popularization of new, non-Aristotelian, that is, non-mechanical philosophical worldviews, leading to the Scientific Revolution, in particular driven by Galileo Galilei's (1564–1642) invention of the pendulum as a viable timekeeping device. Notably, its practical implementation, the pendulum clock, was first achieved by the Dutch polymath Christiaan Huygens (1629–1695) through successful collaboration with his master clockmaker, Salomon Coster (c. 1620–1659).[1] Importantly, mechanical developments were very much at the heart of the Scientific Revolution (Meli 2006). Theoretical knowledge pursued by scholars like Huygens developed in tandem with the practical evolution of knowledge generated by skilled instrument makers (e.g., Klein and Spary 2010). Nevertheless, while the era's leading scholars clearly recognized the importance of the master artisans' practical skills, more often than not they were reluctant to acknowledge their intellectual achievements. In turn, this patronizing attitude may have actually hampered intellectual progress more broadly by excluding some of the most talented and insightful participants from the conversation.

Naturally, the open, tolerant, and transparent conditions in Huygens' seventeenth-century Dutch Republic allowed the nation's leading intellectuals to play a key role before and during the 'Scientific Renaissance.' In an era of increasing maritime trade, the ability to reliably determine one's position at sea was particularly important for the crews of the large fleets of merchant ships hailing from Europe's maritime

---

[1] See de Grijs (2017, Ch. 3) for a recent review of early developments from Galileo's Venice to the Dutch Republic, eventually leading to Huygens' first practical pendulum clock design.

nations.

Yet, accurate and precise maps and charts were closely guarded state secrets, accessible only to the most senior officials; generally available sources of geographical information were of mixed reliability. These problems were compounded by the absence of a reliable means of longitude determination at sea, which required timekeeping to unprecedented accuracy. Consequently, multiple European governments announced rich rewards to anyone who could solve the intractable 'longitude problem' (de Grijs 2017, pp 3-19–3-24).

Huygens had become a major contributor to attempts by the leading European intellectuals to develop a timepiece suitable for navigational purposes on the open seas, away from the relative safety of coastal waters (de Grijs 2017, Ch. 3–6). Sometime during the late 1660s or early 1670s, Huygens had realized that springs could be implemented as highly regular oscillators to drive his clocks, allowing one to achieve much better performance than the usual weight-driven pendulums. Huygens was indeed a forerunner of his time, although his approach to the scholarly enterprise was still predominantly theoretical. At heart, Huygens was an applied mathematician extraordinaire, a scholar equipped with a brilliant mind but lacking the specialized ability needed to put his ideas into practice. To his tremendous credit, he sought the practical skills he needed by engaging the celebrated instrument makers of his time, although he clearly felt that the intellectual aspects of these new developments were his, and his alone. Huygens commissioned a spring-driven watch from the distinguished Parisian clockmaker Isaac Thuret (c. 1630–1706), *Horloger Ordinaire du Roi* (that is, clockmaker to the French Royal court), clockmaker of the *Académie Royale des Sciences* (the Royal Academy of Sciences), and—following its completion in 1671—also of its astronomical observatory (Huygens 1666–1695).[2]

Much earlier, during his first extended sojourn to Paris in the Fall of 1655, Huygens had likely had ample opportunity to meet Thuret, who was then already rapidly developing into France's greatest seventeenth-century clockmaker. The seeds for one of the most fertile mathematician–instrument maker partnerships of the second half of the seventeenth century may well have been planted during this period. Although Huygens' correspondence prior to 1662 does not refer explicitly to Thuret, this could be explained easily by lost correspondence, by Huygens' uncompromising attitude that Thuret did not have anything to do with his scientific inventions—which he considered his own intellectual property (as I will discuss below)—or by the need for secrecy. After all, Huygens was often in direct competition with the Parisian clockmakers, and intellectual piracy ('plagiarism') was rife.[3] In fact, competition with French clockmakers dominates part of Huygens' surviving correspondence. For instance, in a letter dated April 12, 1662, to his Paris-domiciled brother Lodewijk, he pointedly asked,

---

[2] See the *Comptes des Bâtiments du Roi, sous le règne de Louis XIV (1664–1687)*. 1881. Paris: J. Guiffrey, p. 230.

[3] However, note that Chapelain comments in a letter to Huygens of August 20, 1659, about a Parisian clockmaker ("*de notre*") "*who has tried to rob you of your claim.*" The editors of the *Oeuvres Complètes de Christiaan Huygens* (HOC) suggest that this may refer to Isaac Thuret, although without any supporting evidence; 1659-08-20: Chapelain, Jean – Huygens, Christiaan; HOC II:467–469 (No. 655).

> *"... how are these Thuret clocks made, for which my father pays 10 or 12 pistoles* [gold coins] *and [which he] prefers to his own? If we could know the form it could be used to instruct the clockmakers here ..."*[4]

Here, Huygens' query refers to a specific type of clock made by Thuret, with which the scholar and his colleagues in the Dutch Republic were in direct competition. Nevertheless, competition and collaboration with French clockmakers seem to have occurred simultaneously. Clearly, highly promising cutting-edge technical and intellectual developments were happening across northern Europe at the time, largely triggered by the intense competition for the sovereign's 'privileges.'[5] Crucially, however, the Dutch and French competitors needed to build on each other's ideas to make sustained progress. Following a series of letters among Huygens' family members in 1663 and 1664 which involved sending damaged clocks for repair to Thuret (Plomp 1999), the following year Jean Chapelain (1595–1674)—founding member of the French Academy—wrote to Huygens in relation to a French patent the Dutch scholar had been awarded,

> *"… that excellent clockmaker Monsieur Thuret, of whom you yourself have told me much good, visited me yesterday and asked me to offer you his services for the construction of clocks to be used on ships and for their sale and distribution."*[6]

Huygens agreed to employ the Parisian clockmaker,[7] reassured by Chapelain's recommendation that he would be served by Thuret *"incomparably better and with more capacity ... and intelligence than by any other."*[8] Chapelain clearly valued Thuret's intellectual capabilities highly, which should have triggered Huygens' keen interest from the outset. However, despite this praise of Thuret's skills and the ensuing, successful Huygens–Thuret collaboration, the accuracy of Huygens' timepieces still required significant improvements before they could be used reliably for longitude determination.[9]

*Early days*

Already during these early years of their cooperation, Thuret's intellectual creativity had become apparent on a number of occasions, thus showing that the clockmaker was capable of high-level abstract, intellectual contributions. As a case in point, to increase the accuracy of his clocks, he had designed an early version of a remontoire, a device used to automatically rewind a clock at regular intervals without the need to interrupt its timekeeping function.[10] Chapelain informed Huygens that Thuret had noticed that the Dutch scholar's 'secret' was rather similar to his own (a simple maintaining system now known as Huygens' 'endless chain'),[11] except that

> "… *the little chains in your construction testify to a less simple skill than in his own and more subject to arrest as has happened with the one of Monsieur [Pierre de] Carcavi and the one of Monsieur [Henri Louis Habert] de Montmor.*"[12]

---

[4] 1662-04-12: Huygens, Christiaan – Huygens, Lodewijk; HOC IV:109–110 (No. 1004). Unless otherwise indicated, all translations are the author's own.

[5] 'Privileges' only roughly correspond to the patents we are familiar with today (Biagioli 2006). I will nevertheless use the term 'patent' throughout this essay, but one should keep in mind that in the seventeenth century the concept 'privilege' included both 'copyright' and 'patent' in the modern sense.

[6] 1665-03-13: Chapelain, Jean – Huygens, Christiaan; HOC V:267–268 (No. 1352).

[7] 1665-03-26: Huygens, Christiaan – Chapelain, Jean; HOC V:281 (No. 1361).

[8] 1665-04-24: Chapelain, Jean – Huygens, Christiaan; HOC V:340–342 (No. 1398); 1665-06-07: Chapelain, Jean – Huygens, Christiaan; HOC V:370–372 (No. 1417).

[9] It would take until December 31, 1682, before the Dutch East India Company (VOC) defined the requisite accuracy any clock had to achieve before it would be considered successful at longitude determination at sea. The initial resolution called for an accuracy of better than 1 second deviation per 24 hours, although a new VOC resolution dated April 28, 1684, relaxed this to a performance requirement of better than 2 seconds per 24 hours (de Grijs 2017, p. 5-48).

Huygens, instead, downplayed the clockmaker's intellectual contribution. Indeed, in a letter to his father dated February 19, 1665, Huygens sneered that

> "*[Salomon] Coster and [Claude] Pascal made such [a device] a long time ago. The device in my timepiece is intended to rewind it every ¼ minute or so, and the pendulum will perform very regular swings. Do not tell anyone.*"[13]

When Huygens moved to Paris in 1666, he most likely continued to employ Thuret. Meanwhile, the clockmaker continued to hone his formidable technical skills and improve the performance of his timepieces, as we can readily deduce from Jean Richer's (c. 1630–1686) report of his 1672–1673 scientific expedition to the Americas. Richer extensively praised the clocks taken along on the voyage; they had been made by Thuret, "*who by his exactness and the delicacy of his works has surpassed up to now everyone involved in the production of watches and pendulum clocks*" (Huygens 1666–1673).

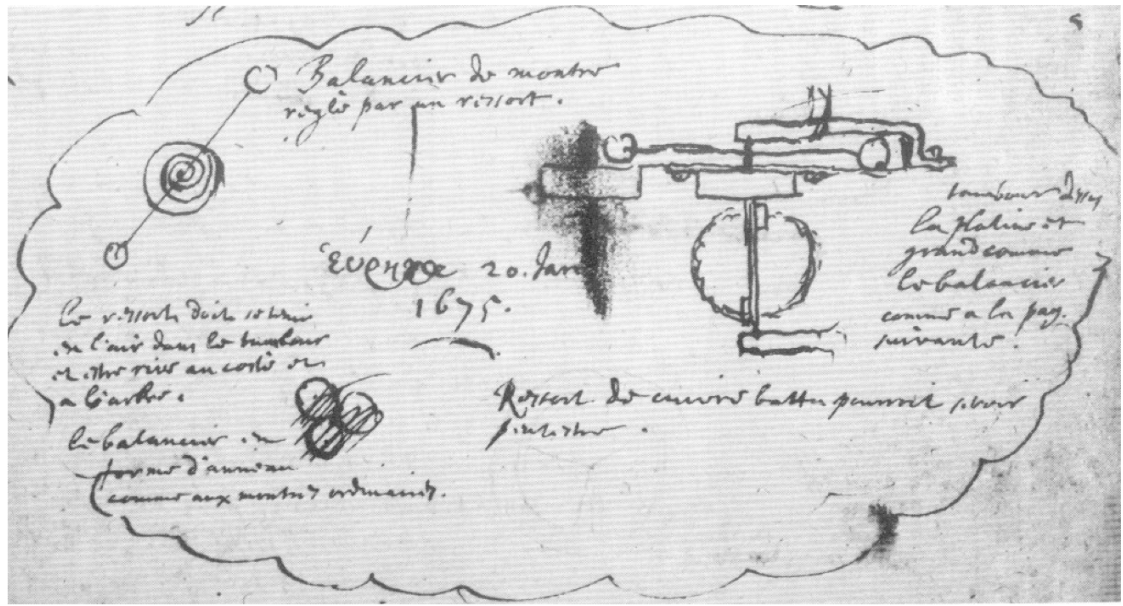

**Figure 1**: Huygens' drawings of a spring balance and escapement; January 20, 1675. (*HOC VII:408; Adversaria, Manuscript E; Leiden University Library, HUG 9, f.18r; reproduced with permission.*)

---

[10] 1665-02-19: Huygens, Christiaan – Huygens, Constantijn Sr.; HOC V:240 (No. 1331).
[11] Huygens invented an early version of this mechanism, known as a 'maintaining-power' or 'going barrel,' around 1658 (based on its publication in Huygens' *Horologium*, 1658).
[12] 1665-10-23: Chapelain, Jean – Huygens, Christiaan; HOC V:510–511 (No 1485).
[13] 1665-02-19: Huygens, Christiaan – Huygens, Constantijn Sr.; HOC V:240 (No. 1331).

Despite their intense collaboration in the 1660s, it took another decade for a first sketch of a new and highly promising design of a spring-driven longitude timepiece to mature; the date of January 20, 1675 prominently accompanies Huygens' novel design ideas (Huygens 1675–1676): see Figure 1. It has been suggested, however, that Huygens may have produced the sketch bearing this date only when he felt the acute need to establish ownership of his invention in the face of a developing, highly acrimonious dispute with Thuret, shortly afterward (Mahoney 2004). Indeed, why would Huygens have added a date to his early sketches? Scientific and technological developments usually occur in fits and starts, and 'eureka moments' are extremely rare.

In any case, this early sketch is accompanied by a set of notes. Some of these do not describe the sketches as such but instead refer to changes required to make the design work. Indeed, Leopold (1980) has suggested that these notes are akin to a running commentary of the discussion that likely took place, when both men must have bounced ideas off each other to arrive at a practically viable design. It is therefore not unlikely that Huygens left Thuret's workshop, went home, and made the sketch at that point, possibly based on an initial model the clockmaker may have produced. If so, this then raises an important question regarding the idea's provenance; indeed, it suggests that Huygens' claim of the design's ownership as its sole inventor, that is, of its intellectual property, may have been altogether spurious.

Be that as it may, Huygens was obviously keen to convert his sketch into an operational timekeeper equipped with a novel spring balance. However, his design as sketched was not good enough for construction of a practical timepiece. His escapement in particular design needed substantial, expert improvements (Leopold 1980); he thus clearly needed access to a skilled instrument maker. Early on January 21, 1675, he first showed his design to Pierre Perrault (c. 1608/11–1680), the Receiver General of Finances for Paris-turned-hydrologist. He then repeatedly but unsuccessfully tried to find Thuret in his workshop, eventually succeeding around noon on January 22. Huygens' enthusiasm received a similarly excited response from the clockmaker.

### *Developing acrimony*

The *Codex Hugenianus E* includes a detailed chronological account—although only from Huygens' perspective—of the controversy that developed (Huygens 1675), on which I will base the discussion here. Their initial meeting prompted Thuret to construct a model watch based on Huygens' design. He then continued to work on an improved version of a spring-driven watch of his own design, requesting that Huygens keep these developments secret.

Huygens was not only the leading scientist of his era, he also had keen business instincts. He recognized that Thuret's new spring-driven watch design (Figure 1) was eminently suitable for a patent application. Therefore, the Dutch scholar approached Jean-Baptiste Colbert (1619–1683), King Louis XIV's right-hand adviser, on January 31, 1675, and showed the minister 'his' new invention. In response, Colbert promised the scholar a French patent, a *privilège du roi*. Thuret indicated that he was keen to be included as co-inventor. To his disappointment, however, Huygens refused, although he sweetened his message by emphasizing that Thuret would profit most from the invention and that his practical skills would always be recognized. Huygens' refusal to assign at least some of the intellectual ownership of their design to Thuret should not be seen as an isolated incident. With hindsight, it is clear that the Dutch scholar did not always fully appreciate the contributions by the instrument makers to the conversion of his designs from ideas to practically viable specimens; he was hence often too quick in interpreting the genuine desire of his skilled technical collaborators to be recognized as infringements of his own intellectual property.

Initially, Huygens agreed to delay their patent application until Thuret had completed a watch model for presentation to the French King. In the meantime, however, Thuret repeatedly announced publicly the implied significance of his own intellectual contributions to the new design, to Huygens' significant displeasure and chagrin. This prompted the latter to speed up his patent application.

Huygens and Thuret did not meet again until February 4, 1675, although their newly acrimonious relationship only came to a head on February 8, when Huygens discovered "*that Thuret had shown my invention to Mr Colbert eight days before me,*" that is, already on January 23, 1675. Huygens described the altercation in detail in his workbook (Huygens 1675; transl. Andriesse 2005, pp 279–280):

> "*Friday 8 [February 1675] a response from Colbert, who had placed his [consent] in the margin. I discussed the matter with the King. Mr [Guichard Joseph] Duverney came to me and said that he understood Thuret to have paid a visit to Colbert and to have submitted to him a petition. I said this to Mr [Claude] Perrault, who told me what he knew up until then: that Thuret had shown my invention to Colbert eight days before I did. I believe that it is true, and that it was the model he showed to me on January 23. And I believe that Mr [Jean] Gallois will have been present. I recall that Thuret displayed some confusion when I told him that I had showed my model to Mr Colbert just shortly before.*"

When confronted by Huygens on February 9, Thuret initially denied the accusation, but he later conceded (Huygens 1675; transl. Andriesse 2005, p 280):

> "*On Saturday the 9$^{th}$ I was at the house of [Charles] Perrault, the controller, who requested that Mr Gallois be present, and at the same time asked that Thuret be fetched. When they arrived, I presented to Thuret my arguments [to then think] that he had, the day after I asked him to make my model, without my knowledge, shown to Colbert just such a model and said that he had invented it. (He denied it.)*
>
> *On the 11$^{th}$ Thuret came to speak with the controller. He admitted what he the day before so strenuously had denied.*"

Is it a coincidence that, being domiciled in Roman Catholic France, Thuret decided to come clean and atone of his sins on a Sunday? We will never know. Huygens added a note in the margin of Thuret's admission of his breach of their mutually agreed secrecy, stating that he had decided to collaborate with the Parisian clockmaker Antoine Gaudron (1640–1714) instead: "*I gave the manufacture of the clock to Gaudron, a cousin of Papin.*"

There are no surviving accounts that can independently verify the extent of Thuret's improvements to Huygens' original design (see also Leopold 1980). However, it is clear that some of the members in their mutual social circles valued Thuret's intellectual contributions more highly than Huygens did. Nevertheless, Huygens could not be swayed.

This acrimonious episode raises more far-reaching questions, however. All available historical evidence suggests that the construction of a practically viable spring-driven timepiece resulting from the Huygens–Thuret collaboration was not the attainment of either man on his own. Both antagonists seem to have contributed their fair share, so that in hindsight Huygens' dismissive stance appears unwarranted. After all, a novel design is not equivalent to an invention if the resulting mechanism does not work. This then raises the question as to whether the entire idea that eventually yielded a working prototype belonged only to Huygens or whether Thuret could also rightly claim a significant intellectual contribution.

Huygens was adamant, however; he clearly felt that Thuret was not his intellectual equal. He repeatedly pointed out that the clockmaker apparently did not fully understand his novel design, at least not at first: *"In explaining it to him, he said (as yet barely understanding it), 'I find that so beautiful that I still can't believe it is so'"* (Huygens 1675–1676; transl. Mahoney 2004). However, Thuret had expressed concern about the practical operation of Huygens' design, particularly as regards the regularity of the resulting oscillations, which he said he had been thinking about himself already for some time (Mahoney 2004). Huygens clearly did not take these defensive comments seriously; he later said that he responded,[14]

> *"… that what he said of the trouble with these vibrations was something contrived to make it appear that he knew something about the application of the spring, but that this itself showed that he had known nothing about it, because, if he had thought about attaching the spring by its two ends,[15] he would have also easily seen that these vibrations were of no concern, occurring only when one knocked or beat against the clock and even then not undercutting the effect of the spring."*

Mahoney (2004) has pointed out that Thuret may, in fact, have been working on a similar design of his own,[16] around the same time, but without recognizing Huygens' brilliant practical solution.

An attempt was made to reach mutually acceptable closure to the tense situation that had developed. On February 25, 1675, Thuret announced that the watch he had been working on was almost finished, and that he was prepared to present it to Huygens shortly afterward. However, Huygens was only willing to accept the timepiece in return for payment, thus publicly showing that the scholar considered the invention his own and that Thuret had merely lent his skills to realize the practical implementation. With the benefit of hindsight, it appears that Huygens was fundamentally mistaken in his assessment of the clockmaker's contributions. Perhaps this reveals his rather volatile temperament, but more importantly he seems to have violated the basic tenet of what we now understand as intellectual property disputes. I will return to the concept of intellectual property rights and their development in Renaissance Europe below.

---

[14] Incorrectly referenced in Mahoney's Note 41 (transl. Mahoney 2004).

[15] The essence of Huygens' novel design was a spring that was affixed *at both ends*, but which could move freely in between.

[16] After all, Thuret clearly understood Huygens' ideas and suggestions sufficiently well to engage in in-depth discussions that led to further improvements. Within hours of Huygens' departure, he made a working model—either from memory or perhaps from an unknown sketch he may have made himself. His material contributions to the final, miniaturized design which Huygens proceeded to show off to all and sundry seem to underscore the validity of Thuret's claim of part-ownership of the underlying development and, indeed, of full credit for its construction.

Despite these morally questionable developments, on September 10, 1675, Thuret was forced to sign a letter in which he gave up any intellectual claims to the spring-driven watch invention.[17] It is likely that the clockmaker agreed to this humiliating defeat for commercial reasons: on February 15, 1675, Huygens had acquired the exclusive right[18] "*to have made watches and clocks of a new invention*" in France for a period of twenty years, specifically for watches consisting of "*a spring shaped as a spiral which regulates the rotations of a free balance, larger and heavier than ordinary specimens.*" Having signed this infamous letter relinquishing all intellectual rights to Huygens, Thuret was henceforth free to continue his engagement as Huygens' preferred craftsman.[19]

*Business as usual*

Huygens' account of the events that occurred between himself and Thuret between January 21 and February 25, 1675, was apparently composed at a later date, and hence it must be considered at least somewhat one-sided.[20] One should also keep in mind that the entire narrative is only known from Huygens' perspective; it may not, in fact, reflect the unbiased truth (Leopold 1980; Plomp 1999, his footnote 43). Nevertheless, a series of surviving letters between Huygens and Henry Oldenburg at the Royal Society in London shed some independent light on the priority of invention. In a letter dated January 30, 1675, Huygens stated,

---

[17] 1675-09-10: Thuret, Isaac – Huygens, Christiaan; HOC VII:499 (No. 2055).
[18] 1675-02-15: Colbert, Jean-Baptiste – Huygens, Christiaan; HOC VII:419–420 (No. 2011).
[19] 1675-09: Huygens, Christiaan – Perrault, Claude; HOC VII:497–498 (No. 2054).
[20] Although I have based this narrative on Huygens' *Oeuvres Complètes*, which is considered an incomplete edition of Huygens' work from a modern point of view, the timeline of events described here follows Huygens' own perception of developments. As such, the *Oeuvres* comprise an adequate representation of the original document, contained in the *Codex Hugenianus E*. However, since the latter is not readily available for further scholarly perusal, while the former collection is, I prefer referencing manuscripts that can be accessed easily for follow-up study.

> *"About Pendulums, I shall tell you ... that I have recently come upon a new invention pertaining to Clocks which I am having [someone] presently working on and which appears to be successful. I include the secret here, in anagram form; as you know, I have done [this] before with new discoveries and for the same reason."*[21]

The anagram[22] read as follows:

```
4 1 3 5 3 7 3 1 2 3 4 3 2 4 2
A B C E F I L M n O r S t u x
```

Rearranged to its original Latin, this becomes *Axis circuli mobilis affixus in centro volutae ferreae* (*The axis of a moving circle attached to the center of an iron coil*); the 'moving circle' is the clock's balance wheel, which drives its movement. Oldenburg responded on February 12, 1675,[23] noting that their mutual friends in Britain were keen to see the new invention demonstrated in practice. However, by that time the row between Huygens and Thuret had already become common knowledge. This prompted Huygens to respond defensively on February 20, 1675:

> *"In my last letter of January 30, I sent you the secret of a new invention pertaining to clocks, of which you since may have been informed. You will already know what it is, because the secret has not been well preserved here owing to the bad faith of the watchmaker I had commissioned to do the work. Already the day after I had informed him of this invention, allowing him to make a model, he promptly made another and proceeded to showing it off without my knowledge to Mr Colbert and to several other people, saying that he was its creator."*[24]

Huygens clearly felt that he needed to offer his perspective of the unfortunate situation, continuing,

> *"The dishonesty of the craftsman, about which I have spoken to you, caused me a lot of trouble and frustration. But eventually, after I had explained and convinced Mr Colbert of his bad behavior, he has done me justice, and he has given me the King's exclusive right for this invention; upon which my plagiarist, having seen that he had become embroiled in a very serious matter, not knowing what to do, requesting everyone he knew to urge me to forgive his mistake and offer him work as before, promising to testify to everyone that he is in no way part of the claimed invention."*

The conflict about the extent of Thuret's intellectual contribution to the invention of a practical spring-driven clock is often thought to have ended their collaboration.

---

[21] 1675-01-30: Huygens, Christiaan – Oldenburg, Henry; HOC VII:399–400 (No. 2003).

[22] Whereas in today's scientific enterprise one would pursue publication of new insights in scholarly journals, without access to such a route to publication prior to 1665, many seventeenth-century researchers resorted to disseminating their discoveries to their fellow learned men in the form of anagrams. These could not be deciphered without actual knowledge of the discovery, thus establishing priority of discovery (Nielsen 2011).

[23] 1675-02-12: Oldenburg, Henry – Huygens, Christiaan; HOC VII:416–417 (No. 2009).

[24] 1675-02-20: Huygens, Christiaan – Oldenburg, Henry; HOC VII:422–424 (No. 2013).

However, in a letter to his brother Constantijn Huygens Jr. dated August 9, 1675, the scholar writes,

> *"What they said about my plagiarist is true ... but I have deferred to the King's advice, where Mr Colbert has promised to send me a decision that would be equivalent to actually having registered the Privilège du Roi. I will wait and see what the effect of this development will be, and I am determined to obtain a confession from this rogue* [ce coquin] *and get the satisfaction I desire,* or else I will leave everything in this country behind ..."[25]

His father seems to have attempted to defuse Christiaan's anger; in the letter's margin, he added an admonition, "*ne saevi, magne sacerdos,*" in essence stating that adversarial conduct is not good for business. Indeed, despite his hurt feelings, Huygens subsequently mentions Thuret's valued craftsmanship in a letter to Oldenburg of November 21, 1675. He specifically discusses a watch made for the Duke of York:

> *"It is made by Thuret, who makes up to now the best [watches] and with great demand. He is the one who has treated me so badly after I had confided [in] him that invention. But, having retracted in the end by a letter which others had obliged him to write to me, and having come to me to ask pardon, I have no difficulties anymore to employ him."*[26]

Thuret's craftsmanship remained unsurpassed well beyond the clockmaker's lifetime. For instance, in 1754 Johan Lulofs (1711–17568)—Professor of Mathematics and Astronomy at Leiden University in the Dutch Republic—extensively praised the clock he used for his observations of the planet Mercury (Lulofs 1754):

> *"The timepiece that I used was made by Thuret in Paris, under supervision of Mr Huigens, which—given its repeatedly confirmed accuracy—I prefer by far with respect to a newer [model]."*

*Modern perspective*

In hindsight, the Huygens–Thuret spring-driven clock represented the start of truly promising developments pertaining to longitude timepieces. Thuret constructed the first operational spring-driven watch, which—although based on Huygens' theoretical breakthrough of 1675—may well be considered the most important development leading toward the eventual construction of accurate mechanical watches.

Plomp's (1999) discussion of a recently discovered longitude clock attributed to Thuret implies that the Parisian clockmaker was much more closely involved in contemporary efforts to tackle the as yet unyielding longitude problem than is often assumed. Indeed, it has become abundantly clear that material contributions from both clockmakers and scientists, working in tandem, were required before the promising prototype resulting from Huygens' invention and Thuret's craftsmanship was eventually transformed into a timepiece deemed sufficiently accurate for measuring longitude at sea.

---

[25] 1675-08-09: Huygens, Christiaan – Huygens, Constantijn Jr.; HOC VII:483–485 (No. 2045).
[26] 1675-11-21: Huygens, Christiaan – Oldenburg, Henry; HOC VII:542–543 (No. 2078).

This narrative of the Huygens–Thuret controversy, its initial causes, and the subsequent developments highlights the pitfalls associated with assigning intellectual credit. From our current, modern perspective, it appears that Thuret drew the short straw. Indeed, Huygens' name is forever associated with the development of longitude timekeepers in a variety of incarnations, while Thuret's contributions are known only to those scholars whose interests may lead them to explore the history of longitude determination at sea in a more than cursory fashion. Despite his material contributions to resolving one of the most pressing problems of his time, Thuret will thus forever remain the 'invisible technician' to all but the most dedicated historians of science.

Notwithstanding apparent support from powerful friends in his social circle, Thuret was unable to protect his intellectual property in the face of Huygens' vindictive insistence on his intellectual ownership and the priority of invention—indeed, this demonstrates the systemic dilemmas associated with such an inherent power imbalance. Undoubtedly, Thuret was also blessed with a keen business acumen, thus prompting him to relinquish his intellectual rights in return for future commercial gains. The tense situation was not made any easier by Huygens' himself, whom we now know to have had a rather difficult personality. He clearly did not suffer (those he considered) fools easily, and in his letters, he often comes across as angry, arrogant, bitter, and/or obstinate, particularly when he felt cornered or opposed (Bos 1991; de Grijs 2017, p. 3-26).

We may wonder whether Thuret would have fared any better in modern times. As a *Gedankenexperiment*, let us therefore consider his contributions to the design of the first spring-driven watches from an intellectual perspective. The concept of 'intellectual property' as enshrined in modern intellectual property law was all but unknown in the seventeenth century. Neither France, nor Britain, the Italian city states, or the Dutch Republic upheld intellectual property rights; instead, inventors sought privileges,[27] which would in essence provide them with a commercial monopoly for a defined period of time (Biagioli 2006). The practice of granting privileges and licensing also allowed governments to prevent the publication of books as well as maps and charts that were either deemed undesirable for any number of reasons or considered state secrets, by withholding permission ('censorship'; Biagiolo 2006; Hughes 2012).

By all accounts, if the Huygens–Thuret row had played out in modern times, both antagonists would most likely have been granted part-ownership of their joint design. The lack of a successful appeal by Thuret at the height of the controversy may well have had its roots in Huygens' apparent lack of understanding of the clockmaker's intellectual capabilities, in the importance of the artisan's contributions to the final design, and/or in the Dutch scholar's obstinate attitude to any claim of perceived 'infringement' of his own ideas. However, there may have been more to the story that has remained hidden by the mists of time. Christiaan Huygens' father, Constantijn Huygens Sr. (1596–1687), is well-known to have had excellent political connections both in the Dutch Republic[28] and to the French court[29]—and so did Christiaan himself apparently, through his cordial relationship with the King's right-hand adviser, Colbert. May these connections have contributed to the French monarch's rapid and positive assessment of the Dutch scholar's request for a Royal privilege? We will most likely never know.

---

[27] Legally, privileges were expressions of the sovereign's will.

In the current era, intellectual claims related to scholarly contributions are often assessed against the widely adopted recommendations of the International Council of Medical Journal Editors:[30]

- Substantial contributions to the conception or design of the work; or the acquisition, analysis, or interpretation of data for the work; **and**
- Drafting the work or revising it critically for important intellectual content; **and**
- Final approval of the version to be published; **and**
- Agreement to be accountable for all aspects of the work in ensuring that questions related to the accuracy or integrity of any part of the work are appropriately investigated and resolved.

Although Thuret was stopped short in claiming part-ownership of their invention because of an historical accident, I believe that diligent application of all four recommendations would have vindicated his intellectual stature. The 'invisible technician' was indeed short-changed—intellectually, at least.

---

[28] He served as secretary to two of the nascent country's Royal princes.

[29] The French King Louis XIII appointed him as Knight of the Order of Saint-Michel in 1632; a decade later, the French King allowed him to display a golden lily on a blue field in his coat of arms.

[30] *Defining the Role of Authors and Contributors*
http://www.icmje.org/recommendations/browse/roles-and-responsibilities/defining-the-role-of-authors-and-contributors.html [Accessed February 23, 2018]